\documentclass[12pt]{article}

\usepackage[dvips]{graphics}

\begin{document}

\title{Nonlinear optical response of wave packets on quantized potential energy surfaces}
\author{Kunio Ishida and Fumihiko Aiga\\
Corporate Research and Development Center, Toshiba Corporation\\
1 Komukaitoshiba-cho, Saiwai-ku, Kawasaki 212-8582, Japan\\
and\\
Kazuhiko Misawa\\
Department of Applied Physics, \\
Tokyo University of Agriculture and Technology\\
2-24-16 Naka-cho, Koganei, Tokyo 184-8588, Japan
}
\date{}

\maketitle

\begin{abstract}
We calculated the dynamics of nuclear wave packets in coupled
 electron-vibration systems and their nonlinear optical responses.
We found that the quantized nature of the vibrational modes 
 is observed in pump-probe spectra particularly in weakly interacting electron-vibration
 systems such as cyanine dye molecules.
Calculated results based on a harmonic potential model and molecular
 orbital calculations are compared with experimental results, and we
 also found that the materials parameters regarding with the geometrical structure of potential energy surfaces
 are directly determined by accurate measurement of time-resolved spectra.
\end{abstract}

\section{Introduction}
\label{intro}

Optical control of quantum-mechanical states has attracted attention of many authors\cite{review}.
In particular, recent progress in laser technology has made it possible
to synthesize desired optical pulses, and they are expected to be a tool
to control physical properties of various materials, {\it e.g.},
luminescence intensity, lifetime of excited states, or chemical reaction rate.

We, however, have not found effective methods for coherent control of
those properties, since it is impossible to design  appropriate optical
processes without the knowledge of the structure of the electronic states and/or the potential energy
 surfaces (PESs) of materials.
Thus, it is important to determine the values of the material
parameters regarding with the ground-state and/or excited-state PESs to realize coherent control of above-mentioned physical properties.
As a first step toward this approach, we study methods to trace the motion of wave packets on each PES
 after photoexcitation in order to understand the structure of PESs.

Experimental studies on wave packet dynamics have been performed with
ultrashort laser pulses since the pioneering work by Zewail {\it et
al.}\cite{zewail}, and previous studies on molecules in solution give a clue to understand the
 motion of wave packets created by optical pulses\cite{cc1,cc2,cc3,kumar}.
Although the nonlinear response of wave packets on PESs has been
understood by means of the motion of a ``classical point'' on PESs,
our recent experiments on cyanine dye molecules
 have shown that the wave packet dynamics in those
molecules is understood only by taking into account the quantum nature of
the vibrational modes of the molecules\cite{optcom}.
According to our preliminary calculations\cite{optcom}, this particular feature is obtained when the interaction between
electrons and the vibrational modes is weak.
To be more precise, the quantum nature of vibrational modes becomes prominent when the average number of vibrational quanta contained
in each wave packet is $\sim 1$, which shows that the wave packet motion
is restricted in the vicinity of the bottom of the PESs.
In this case, the relaxation energy of the Franck-Condon state is
comparable to the energy of vibrational quanta $\hbar \omega$.

In this paper we calculate the dynamics and the nonlinear optical
response of wave packets on excited-state PES when quantization of vibration modes plays an
important role, {\it i.e.}, in weakly coupled electron-vibration
systems.
In this particular case, we found that the impulse approximation is
effective to analyze the experimental data.
We also show that accurately measured pump-probe signals help us
determine the values
of the material parameters regarding with the structure of PESs, which
are, as mentioned before, important to
design the coherent control methods of wave packets by laser pulses.

The paper is organized as follows: in Section \ref{model} we introduce
the model and its basic properties.
We show the calculated results of the pump-probe signal including the
molecular orbital calculation for cyanine dye molecules in Section
\ref{results}.
Section \ref{discussion} is devoted to discussion and conclusions.

\section{Model}
\label{model}

In order to describe the dynamics of wave
packets in typical organic molecules, we take into account relevant two
electronic states which correspond to the ground and the excited
states and interact with multiple vibrational modes.
Thus, we employ the simplest model suitable for the present purpose,
{\it i.e.}, a two-level model on potential surfaces with $M$
modes described by the following Hamiltonian:
\begin{equation}
{\cal H}   =  \sum_{j=1}^M \left \{ \frac{p_j^2}{2} + \frac{\omega_j^2}{2} q_j^2
 +  \left ( \hbar \omega_j s_j^2  + 
        \sqrt{2 \hbar \omega_j^3} s_j q_j + {t_j \omega_j^2 \over 2} q_j^2
           \right ) |\uparrow \rangle \rangle \langle \langle \uparrow | \right \} 
 +  \hbar \varepsilon |\uparrow \rangle \rangle \langle \langle \uparrow | .
\label{ham0}
\end{equation}
In Eq.\ (\ref{ham0}) $|\uparrow \rangle \rangle$ denotes the excited electronic state
of a single molecule, and its ground electronic state is  $| \downarrow \rangle \rangle$.
$p_j$ and $q_j$ are the normalized momentum and the coordinate of the
vibration modes of the molecule, and $s_j$ is the
dimensionless coupling constant between an electron and the
$j$-th vibrational mode.

Since, in weakly coupled electron-vibration
systems, the wave packet motion is restricted only  in the vicinity
of the bottom of PES as mentioned before, the parabolic approximation
is appropriate in the present study.
In particular, a Gaussian wave packet which is generated by the
vertical transition by optical field does not distort
while it moves on parabolic PESs.
This property appears to be quite useful, as the optical response within the impulse
approximation is obtained analytically as shown in the next section.

In the rest of the paper, we assume that the vibrational
frequencies on both PESs are the same, {\it i.e.}, we take $t_j=0$,
which is appropriate for cyanine dye molecules as shown in the next section.
The details of the quantization of the Hamiltonian (\ref{ham0}) is also described
in the Appendix \ref{eigen}, and
we show in this section only the eigenvalues and eigenstates of Hamiltonian (\ref{ham0}) 
corresponding to the ground-state PES and the excited-state PES
for $t_j=0$:
\begin{enumerate}
\item ground-state PES \\
\begin{eqnarray}
E_{n_1,n_2,...,n_M}^\downarrow & = & \hbar \sum_{j=1}^M n_j \omega_j \\
|n_1,n_2,..., n_M \downarrow \rangle & = &|n_1,n_2,...,n_M \rangle \otimes | \downarrow \rangle \rangle.
\end{eqnarray}
\item excited-state PES\\
\begin{eqnarray}
E_{n_1,n_2,...,n_M}^\uparrow & = & \hbar \left ( \sum_{j=1}^M n_j \omega_j + \varepsilon \right )\\
|n_1,n_2,...,n_M \uparrow \rangle & = & \prod_{j=1}^M
    e^{-s_j(a_j^\dagger-a_j)}| n_1,n_2,...,n_M \rangle \otimes | \uparrow \rangle \rangle,\nonumber \\
\end{eqnarray}
\end{enumerate}
where the boson number states (Fock states) are defined by
\begin{equation}
|n_1,n_2,...,n_M \rangle = \bigotimes_{j=1}^M \frac{(a_j^\dagger)^{n_j}}{\sqrt{n!}} |0 \rangle.
\end{equation}

\section{Calculated results}
\label{results}

\subsection{Pump-probe signals of wave packets on harmonic potential surface}
\label{potcal}

When we consider that the temporal width of both the pump and the probe
pulses are small compared with the period of vibration modes
$2\pi/\omega_j$, we apply the impulse approximation to calculate the
probe signal\cite{mukamel} as
\begin{equation}
A(\Omega;T) \propto {\rm Im} \left [ \int_0^\infty P(T,t) e^{i\Omega (t-T)} dt \right ],
\label{plz}
\end{equation}
where the time-dependent polarization induced by the probe pulse
$P(T,t)$ is given by
\begin{equation}
P(T,t) = {i \over \hbar} \langle  \Phi(T) | [V(t-T),V] | \Phi(T) \rangle \theta (t-T).
\end{equation}
$\Omega$ and $T$ denote the frequency and the delay of the probe pulse,
respectively, and $\theta(t)$ is the step function.

The wavefunction $|\Phi(T) \rangle$ is the Franck-Condon state
multiplied by the time-evolution operator with regard to ${\cal H}$
and is described by
\begin{eqnarray}
|\Phi(T) \rangle & = & e^{-\frac{i{\cal H}T}{\hbar}}|s_1,s_2,..., s_M ; \uparrow \rangle \nonumber \\
& = & e^{-i\varepsilon T}| s_1 e^{-i\omega_1 T},s_2 e^{-i\omega_2  T},... s_M e^{-i\omega_M T}; \uparrow \rangle,\nonumber \\
\label{wf}
\end{eqnarray}
where $|\alpha_1,\alpha_2,... \alpha_M ; \uparrow \rangle$ is a Gaussian wave packet
on the excited-state PES, {\it i.e.}, a coherent state of the vibration
modes defined by
\begin{eqnarray}
& & |\alpha_1,\alpha_2,..., \alpha_M ; \uparrow \rangle  \nonumber \\
& = & \sum_{n_1,n_2,..n_M=0}^\infty \prod_{j=1}^M \frac{e^{-\frac{|\alpha_j|^2}{2}}\alpha_j^{n_j}}{\sqrt{n_j!}}|n_1,n_2,..., n_M \uparrow \rangle.
\end{eqnarray}
The polarization operator is given by $V= |\uparrow \rangle \rangle
\langle \langle \downarrow | + |\downarrow \rangle \rangle \langle
\langle \uparrow |$ and 
its time-dependent form $V(t)$ is
\begin{equation}
V(t) = e^{\frac{i{\cal H}t}{\hbar}} V e^{-\frac{i{\cal H}t}{\hbar}},
\label{dipole}
\end{equation}
in the interaction representation.

After algebraic calculations, $P(T,t)$ is obtained as 
\begin{equation}
P(T,t) = {\rm Im} \left ( \frac{e^{i\varepsilon (t-T)}}{\hbar}\prod_{j=1}^M \bar{p}_j(T,t) \right ) \theta(t-T),
\label{polmodel}
\end{equation}
where
\begin{eqnarray}
\label{polfinal}
\bar{p}_j(T,t)  & = &   \exp\{-s_j^2(1-e^{i\omega_j t} -e^{-i \omega_j T}+ e^{-i\omega_j t} + e^{i \omega_j T} - e^{-i \omega_j (t-T)}) \} \\
& = & e^{-s_j^2(1+2i\sin \omega T)}\sum_{n=0}^\infty \frac{s_j^{2n}}{n!}
\sum_{l+m \leq n} \frac{(-)^m}{l!m!(n-l-m)!}e^{-i\omega_j (n-2l)t}
e^{-i\omega_j (n-l-m)T}.
\label{polpolynomial}
\end{eqnarray}

Combining Eqs.\ (\ref{plz}), (\ref{polmodel}), and
(\ref{polpolynomial}), we found that finite probe signal is obtained
only when the resonance condition
\begin{equation}
\Omega = \varepsilon + \sum_{j=1}^M k_j \omega_j \ \ \ (k_j = 0,\pm 1,
\pm 2,..),
\label{discrete}
\end{equation}
is satisfied.

For $M=1$, $P(t,T)$ has a far simpler form and $A(\Omega;T)$ is
analytically obtained for arbitrary value of $\Omega$ and $T$.
When we consider the case for $M=1$ the indices of the parameters are omitted 
for brevity in the rest of the paper.
The Fourier transformation of $P(t,T)$ is performed in a 
straightforward manner and $A(\Omega;T)$ is given by
\begin{eqnarray}
A(\Omega;T) & \propto & \frac{1}{\hbar} {\rm Im} \left [ e^{-s^2(1-2i\sin \omega T)}\sum_{n=-[\varepsilon \omega]}^\infty \frac{e^{-in \omega
        T}}{(1-e^{-i\omega T})^{n/2}} J_{n} \left (2s^2\sqrt{1-e^{-i\omega
    T}} \right ) \delta (\Omega - \varepsilon - n \omega )
\right . \nonumber \\ 
& + & \left . e^{-s^2(1+2i\sin \omega T)}\sum_{n=-\infty}^{-[\varepsilon \omega]-1} \frac{e^{in\omega
        T}}{(1-e^{i\omega T})^{n/2}} J_{n} \left (2s^2\sqrt{1-e^{i\omega
    T}} \right ) \delta (\Omega + \varepsilon + n \omega )
\right ],
\end{eqnarray}
where $J_n(z)$ is the Bessel function of the first kind and $[a]$
denotes the largest integer which does not exceed the value of $a$.
When $s^2 \gg 1$, vibration eigenstates are densely distributed between the
Franck-Condon state and the bottom of the excited-state PES, which means that
the classical picture based on continuous PESs is recovered.
For example, $A(\Omega; 2n\pi/\omega)\ (n=0,1,2,3...)$ is obtained as
\begin{equation}
A(\Omega; 2n\pi/\omega) \propto \frac{1}{\hbar} \sum_{j=0}^\infty \frac{e^{-s^2} s^{2j}}{j!}
\delta (\Omega - \varepsilon - j\omega).
\label{fcsignal}
\end{equation}
The coefficients on each $\delta$-function in Eq.\ (\ref{fcsignal}) show that the probe
response as a function of probe frequency is given by the Poisson distribution which approaches the
normal distribution in the limit of $s\rightarrow \infty$.

\begin{figure}
\scalebox{0.4}{\includegraphics*{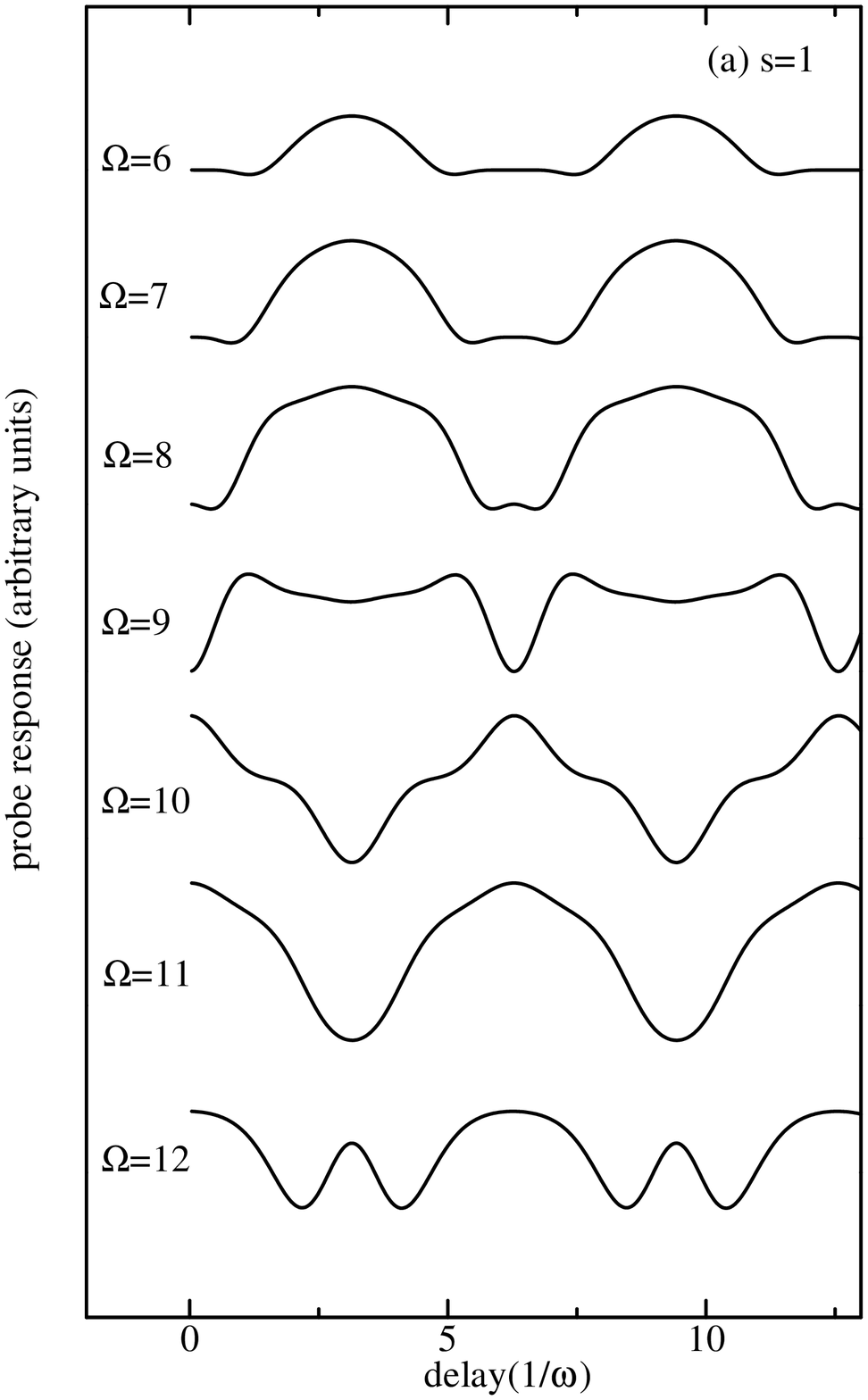}} 
\scalebox{0.4}{\includegraphics*{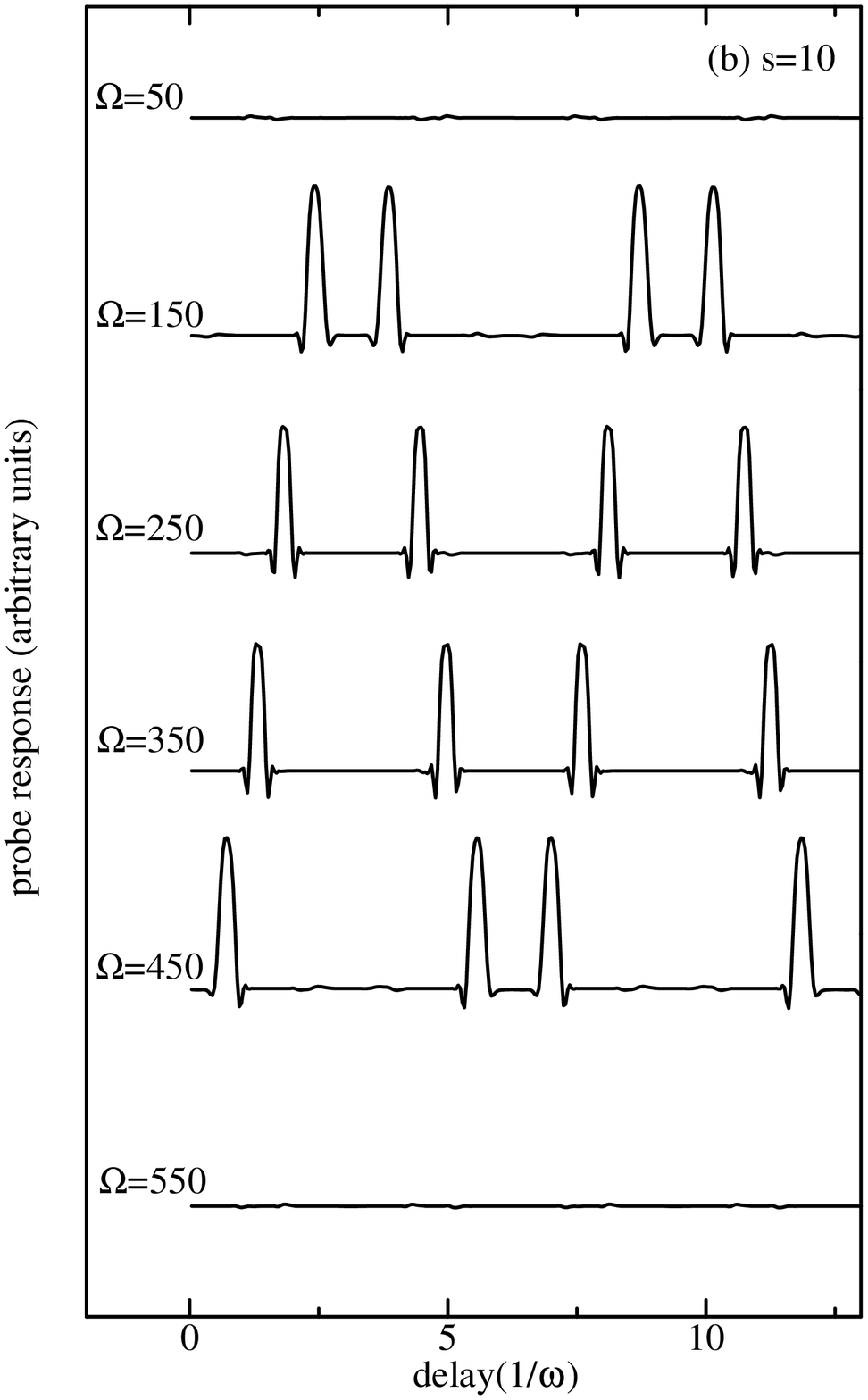}}
\caption{Probe response for $M=1$ and (a) $s=1,\varepsilon=10$, (b)
  $s=10,\varepsilon=100$.
$\Omega$ denotes the probe frequency in units
of $\hbar\omega$.}
\label{signal1}
\end{figure}
We numerically calculated $A(\Omega ; T)$ for arbitrary values of
$\Omega$ or $T$, 
and Figs.\ \ref{signal1}-(a) and (b) show the probe response as a function of $T$ for several values of
$\Omega$ for $s=1$ and $s=10$, respectively.
When $s=10$, the temporal behavior of the probe response is understood
by means of a
classical PES, {\it i.e.},  finite reponse to the probe pulse is observed only when
the center of the wave packet passes the point on the PES at which the
probe pulse resonates to the transition between $|\downarrow \rangle$ and $|\uparrow \rangle$.
However, when $s=1$, the quantum nature of the vibration mode
appears to be important and the shape of the probe spectra as a function
of pump-probe delay $T$ apparently differs from that obtained from
classical PESs, {\it i.e.},
the probe response is finite even when the center of the wave
packet is not on the resonating point of the PES to the probe pulse.
In other words, the probe signal is temporally broadened due to the finite spatial width of the quantum-mechanical wave packet.

Reflecting the motion of the wave packet, the probe signal shows a oscillatory behavior as a
function of $T$.
In the classical limit the peaks of the oscillation in the spectra gradually move as the
probe frequency is changed, whilst the peaks and the valleys in
the spectra suddenly interchanges with each other as the change of the probe
frequency in the quantum-mechanical regime.
Thus the time-resolved spectra translationally move
along the temporal axis as if their phase shifts by $\pi$.
As Eq.\ (\ref{polpolynomial}) shows, however, the phase of the oscillation
as a function of $T$ does not alter, and thus the peak shift in the spectra
cannot be discussed in terms of the phase of the oscillation.
Hence, we stress that time-dependent shift of the phase in each Fourier
component which is often taken into account in experimental studies
is not necessary to analyze the time-resolved spectra to
discuss the wave packet motion.

The quantum nature of the PESs is also shown in observing
the ``turning points'' of the wave packet motion.
When probe frequency is varied, we observe spectral structure which
shows the turning points and they help us determine the values of
material parameters\cite{optcom}.
Since, however, both the average and the
variance of the number of vibrational quanta in a Gaussian wave packet
$|s \uparrow \rangle$ 
are $s^2$, the probe response of wave packets is obtained for higher/lower
probe frequencies than those at the classical turning points.
As a result, the spectra become blurred as shown in experiments\cite{optcom}.
However, the peaks of the spectra still gives the turning points in
classical sense,
and therefore we expect that accurate measurement of time-resolved
spectra is still an effective tool to obtain the values of
 materials parameters.

It has been pointed out that the probe signals include the
optical response of wave packets on the ground-state PES as well as those
on the excited-state PES\cite{cc2}.
Since, as shown in Eq.\ (\ref{plz}), the impulse approximation takes into
account the signals only from the excited-state wave packets, it is
necessary to consider the validity of the impulse approximation for
the probe pulses with finite width.

Using the standard third-order perturbation theory\cite{mukamel}, we
found that the
contribution of ground-state wave packets to the probe response can be
estimated by the transition probability between the ground and excited
PESs,
\begin{equation}
r_{n_1 n_2 ... n_M}^j 
     =  \prod_{j=1}^M \exp \{-2s_j^2 (1-\cos \omega_j \tau)\} \frac{\{2s_j^2(1-\cos \omega_j \tau)\}^{n_j}}{n_j!},
\label{trpre}
\end{equation}
where $\tau$ denotes the temporal width of the pump pulse.
The detail of the calculation to derive Eq.\ (\ref{trpre}) is presented in
Appendix \ref{contr}.

Figure {\ref{matele}} show $r_n$($n=0,1,2,3,4$) for $s=1$ and $s=10$ 
for $M=1$.
\begin{figure}
\scalebox{0.5}{\includegraphics*{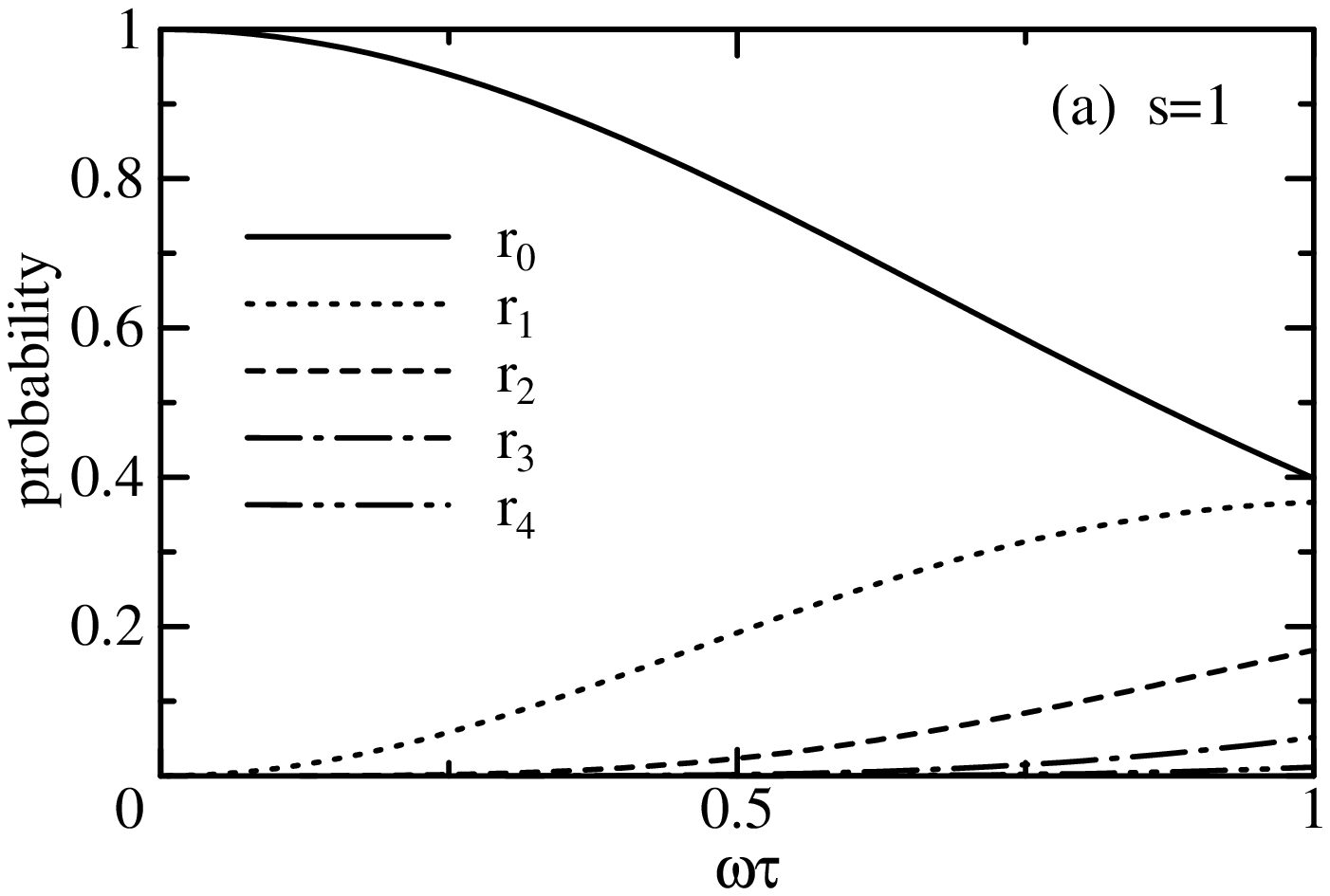}} 
\scalebox{0.5}{\includegraphics*{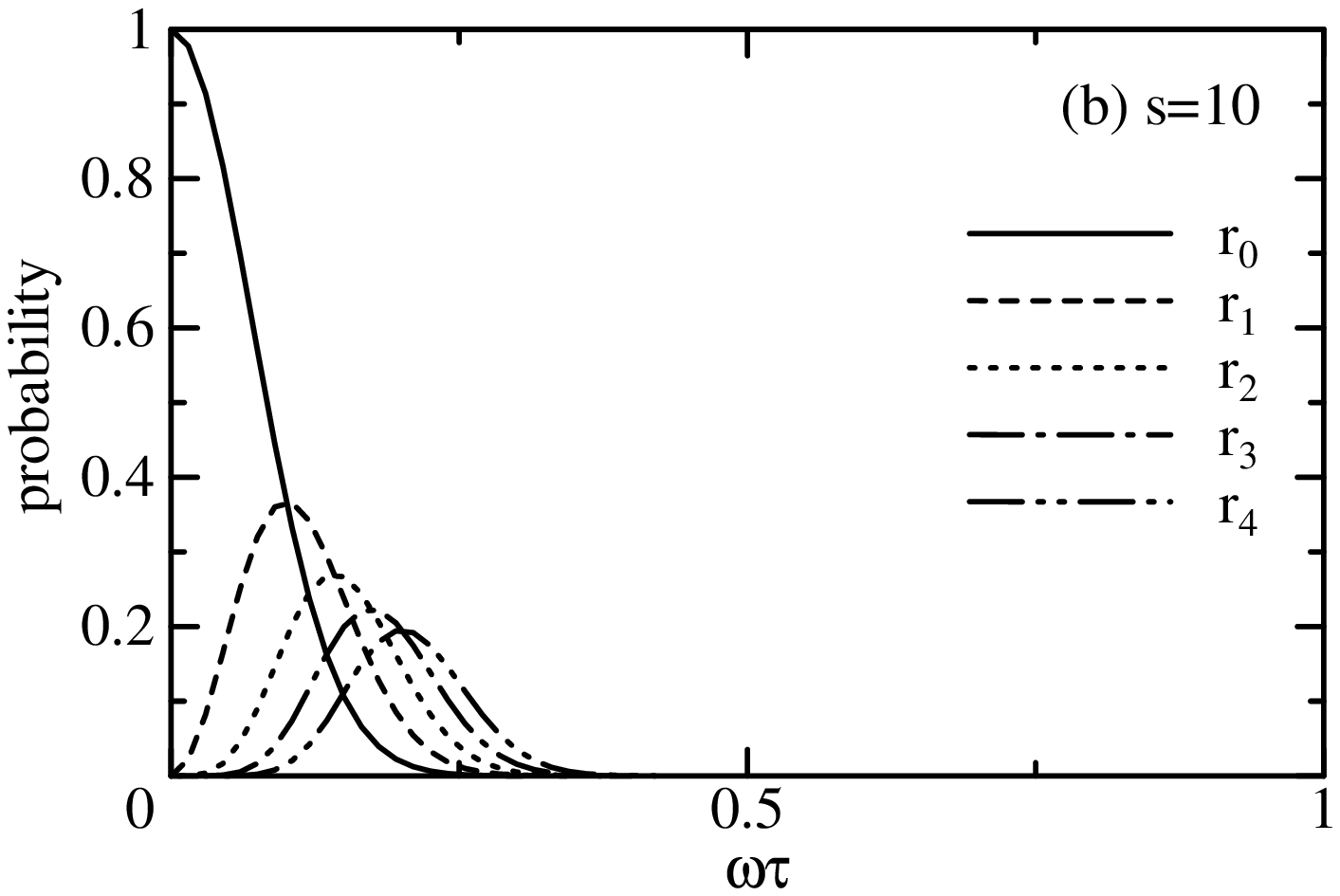}}
\caption{Transition probability $r_n$ ($n=0,1,2,3,4$) for $M=1$ and
  (a)$s=1$, (b)$s=10$.}
\label{matele}
\end{figure}
For $s=1$, $r_n$ for $n \neq 0$ is small compared with $r_0$ for $\omega \tau <
0.5$, which corresponds to the case where the pump pulse is shorter than
the tenth of the vibrational period.
For this short pulse, the ground-state wave packets generated by the pump
pulse are dominated by $|0
\downarrow \rangle$ which does not contribute to the temporal
oscillation of the probe spectra, and hence only the motion of  excited-state 
wave packets is reflected on the time-resolved spectra. 

On the other hand, for $s=10$, $r_n$ for $n \neq 0$ is comparable to
$r_0$ and hence the ground-state wave packets consist of multiple
vibrational states.
In this case, the time-resolved spectra contain oscillatory structure
due to the motion of ground-state wave packets as well as that of the
excited-state wave packets, and the impulse approximation fails.

Hence, we found that the validity of the impulse approximation is estimated by
calculating the values of $r_n$.
Comparing $r_0$ with $r_1$, we found that the impulse approximation works
when $p_c=r_1/r_0 = 2 s^2 (1- \cos \omega \tau)$ or, in general, all
of the corresponding factors $p_c^j$ for $j$-th mode are sufficiently small.
In particular, for weakly coupled electron-vibration systems
such as dye molecules ($s \leq 1$), the probe signals given by the wave packets on the
ground-state PES are independent of the delay $T$ for short probe
pulses used in experiments, and thus the
present calculation is valid to discuss the motion of the excited-state wave packets.

\subsection{Pump-probe response of cyanine dye molecules}

\subsubsection{Molecular Orbital Calculations}
\label{MOcalc}

Before we apply the present theory to the optical properties of DTTCI (diethylthiatricarbocyanine
iodide) measured in our previous study\cite{optcom}, we calculate the electronic
state of DTTCI by the molecular orbital calculations.
DTTCI is dissociated into ${\rm C_{25}H_{25}N_{2}S_{2}^+}$
and ${\rm I^-}$ in the ethanol solution.
Therefore, we consider ${\rm DTTC^+}$(${\rm C_{25}H_{25}N_{2}S_{2}^+}$)
as a model molecule, where the solvent effect is not considered.
The optimized geometry for the singlet ground state of ${\rm DTTC^+}$ was obtained
in the restricted Hartree-Fock (RHF) approximation, and the
vibrational frequency analysis was performed.
The optimized geometry for the first singlet excited state was obtained
in the singly excited configuration interaction (CIS) approximation, and the
vibrational frequency analysis was performed.
The basis set employed in this work was 6-31G(d)\cite{G98}, and
all calculations were performed with the Gaussian 98 program
package\cite{G98}.

\begin{figure}[h]
\scalebox{0.5}{\includegraphics*{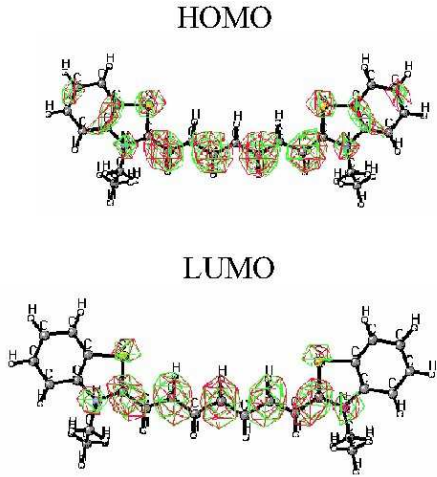}}
\caption{The probability amplitudes for HOMO and LUMO of ${\rm DTTC^+}$.}
\label{prob}
\end{figure}

The probability amplitudes for the highest occupied molecular orbital (HOMO)
and the lowest unoccupied molecular orbital (LUMO) of DTTC$^+$ are given in Fig.\
\ref{prob}.
The character of the first excited state in the CIS calculation
is the excitation from HOMO to LUMO.
The large oscillator strength for this excitation, 2.94 in the CIS calculation,
indicates that DTTCI is the efficient dye molecule.

\begin{figure}
\scalebox{0.5}{\includegraphics*{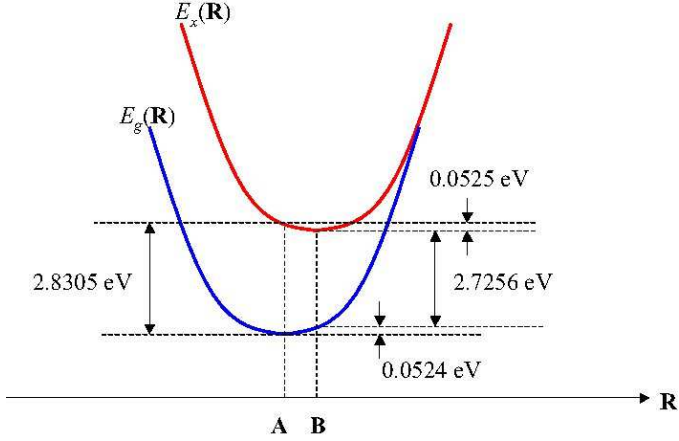}}
\caption{Schematic representation of the potential energy surfaces
for the ground and the first excited states of ${\rm DTTC^+}$.}
\label{sche}
\end{figure}
Schematic representation of the potential energy surfaces
for the ground $(E_g({\bf R}))$ and the first excited states $(E_x({\bf R}))$
of ${\rm DTTC^+}$ is illustrated in Fig.\ \ref{sche}.
The minimum energy configuration for the ground state is shown as $\bf{A}$, and
that for the first excited state is shown as ${\bf B}$.
The difference of the configuration between $\bf{A}$ and $\bf{B}$ is quite small,
and the energy difference $E_x({\bf A})-E_x({\bf B})$ is so small as 52.5 meV.
The difference between the vibrational frequency for the 
first excited state and that for the ground state is also quite small.
Therefore, the weak coupling model can be applied to this problem, and
the vibrational modes for the first excited state can be considered to be the same
as those for the ground state.

In the Franck-Condon picture, absorbing the pump light,
the ${\rm DTTC^+}$ molecule is vertically
excited from the ground state to the first excited state,
and the ${\rm DTTC^+}$ molecule
is placed on the configuration ${\bf A}$ on the potential energy surface of the first
excited state $(E_x({\bf A}))$.
In order to investigate the time-dependent nuclear dynamics of ${\rm DTTC^+}$ molecule
after the excitation, molecular dynamics (MD) calculation is necessary.
Considering the nucleus as classical particles, the MD of ${\rm DTTC^+}$ can be described
by the Newtonian equation of motion with the potential $(E_x({\bf R}))$.
Although we have not performed the full MD calculation,
we have calculated the force at $(E_x({\bf A}))$,
\begin{equation}  
{\bf F}=\frac{\partial E_x({\bf R})}{\partial {\bf R}}\Bigg|_{{\bf R}={\bf A}} \ ,
\end{equation}  
as the measure of the initial dynamics.
The vector ${\bf F}$ has 162 components corresponding to the $x$, $y$, and $z$ directions
of each atom.
Then, the scalar product of {\bf F} with each unit displacement vector
${\bf u}_i (i=1 \sim 156)$ corresponding to each harmonic vibration on $(E_x({\bf R}))$
has been calculated.
The vector ${\bf u}_i$ has 162 components as well.

\begin{table}[htbp]
\begin{center}
\begin{tabular}{|r|r|r|}
\hline
mode & frequency(cm$^{-1}$) & {\bf F}$\cdot$ {\bf u}$_i$ \\
\hline
1 & 15.0427 & 0.06 \\
8 & 86.5312 & -0.06 \\
15 & 191.0350 & -0.07 \\
17 & 219.8474 & -0.06 \\
22 & 292.0076 & -0.12  \\
31 & 443.1244 & -0.07  \\
37 & 550.7053 & -0.06 \\
39 & 574.0371 & 0.09 \\
45 & 645.1332 & -0.19 \\
50 & 784.4855 & 0.06 \\
51 & 792.6894 & 0.09 \\
52 & 800.6848 & 0.07 \\
53 & 800.7149 & 0.10 \\
84 & 1232.1675 & -0.05 \\
96 & 1373.7985 & 0.05 \\
110 & 1544.6282 & 0.08 \\
114 & 1625.5083 & -0.05 \\
126 & 1753.8693 & 0.11 \\
128 & 1776.6010 & 0.05 \\
129 & 1784.0095 & 0.11 \\
130 & 1785.3942 & 0.08 \\
131 & 1827.9013 & 0.11 \\
\hline
\end{tabular}
\caption{The vibrational frequency modes on Ex({\bf R}), having larger absolute value for the scalar product with {\bf F}.}
\label{freq}
\end{center}
\end{table}
The vibrational frequency modes, having larger absolute value for the scalar product with ${\bf F}$,
are given in Table \ref{freq}, where ${\bf F}$ is normalized.
The mode 45 has the largest absolute value for the scalar product.
Considering the beat of our experiment has the period of $\sim$ 230 fsec,
the frequencies for mode 15 is close to the experimental frequency.
For further verification of the experiment, we have to perform the MD calculation .

\subsubsection{Pump-probe spectra of DTTCI}

To compare with the experimentally observed pump-probe spectra\cite{optcom},
we calculated the probe response based on Eq.\ (\ref{polmodel}) for $M=2$.
The values of the parameters of the vibrational frequencies are
$\hbar \omega_1=18$meV (mode 1) and $\hbar \omega_2=58$meV (mode 2).
The electron-vibration coupling constant for the first mode $s_1$ is fixed
to 1.
These values are taken from Ref.\ \cite{optcom}.

\begin{figure}
\begin{center}
\begin{tabular}{c}
\scalebox{0.77}{\includegraphics*{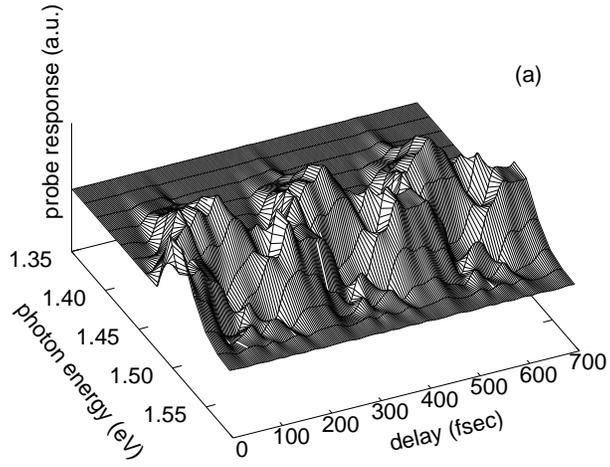}}\\
\scalebox{0.77}{\includegraphics*{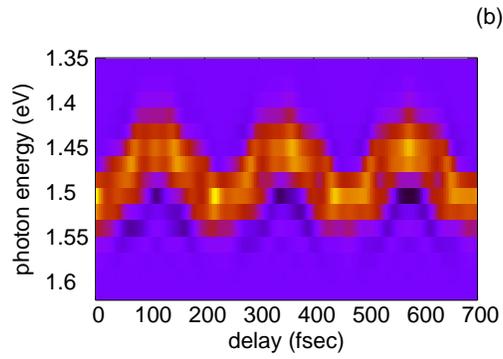}}
\end{tabular}
\caption{Calculated pump-probe spectra for DTTCI as a function of probe
  photon energy and
 the pump-probe delay time: (a) bird's-eye view (b) palette-mapped plot. The values of the parameters are: $\hbar
 \omega_1=18$meV, $\hbar \omega_2 = 58$meV, $\hbar
 \varepsilon=1.5$eV, $s_1=1$, and $s_2=0.6$.}
\label{spectra}
\end{center}
\end{figure}
The calculated results of the pump-probe spectra for $s_2=0.6$ is shown in
Figs.\ \ref{spectra}-(a) and (b).
Although probe response is finite only for discrete values of
probe frequencies (Eq.\ \ref{discrete}), we draw lines which connect
spectral lines with finite response, so as to make the overall behavior of spectra
clearly shown.

The motion of the excited-state wave packet is understood by tracing the
peak of the spectra.
However, as shown in Fig.\ \ref{spectra}-(b), the trajectory of the peak
of the spectra corresponds to that for the vibration mode 1,
although fine structure due to the mode 2 is also seen in the spectra
(see Fig.\ \ref{spectra}-(a)).
In fact, the amplitude of the oscillation in Fig.\ \ref{spectra}-(b)
is $\sim$ 70meV which corresponds to the distance between the two
turning points for the mode 1 ($4s_1^2 \hbar \omega_1=72$meV).

As the value of $s_i$ decreases, the trace of the peak of the spectra
is blurred and the turning points become unclear as we mentioned in Section \ref{potcal}, and 
for this reason the turning points corresponding to the mode 2 
are not seen in the spectra.
As Fig.\ \ref{spectra}-(a) shows, however, the effect of the mode 2 is
clearly
seen in the fine structure of the spectra, which means that we understand
the wave packet motion corresponding to the mode 2 by accurate measurement of
time-resolved spectra.
To be more precise, we found that the shape of the spectra is sensitive to the value of
$s_2$ in the present case and that the $s_2$ is determined by careful
observation of the spectral shape.
\begin{figure}
\scalebox{0.5}{\includegraphics*{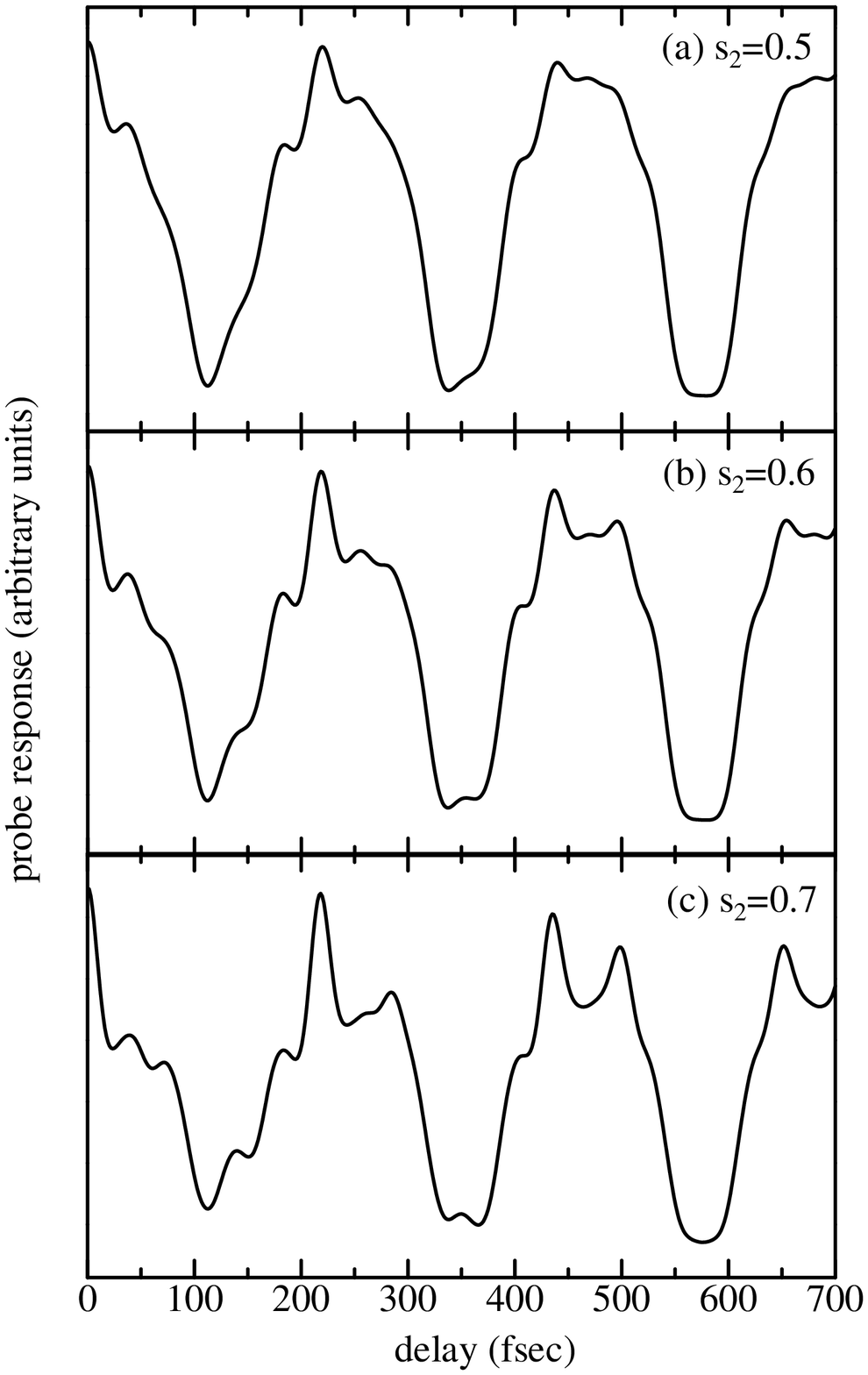}}
\caption{Calculated probe response at $\Omega=1.52$eV for
$\hbar \omega_1=18$meV, $\hbar \omega_2=58$meV, $s_1=1$ and
(a)$s_2=0.5$, (b)$s_2=0.6$, (c)$s_2=0.7$.}
\label{specdttci}
\end{figure}
In Figs.\ \ref{specdttci}-(a)-(c) we show the results for $s_2=0.5$,
0.6, and 0.7 with the values of the other parameters fixed to those in
Fig.\ \ref{spectra}.
The figures show that the shape of the spectra changes rapidly as
$s_2$ is varied.
Thus, we can determine the materials parameters regarding with
weakly coupled mode by a sort of pattern matching methods for time-resolved spectra,
though the precise methods to determine these values have not been well-established.
In the case of DTTCI, we found that $s_2=0.6$ is most feasible by
using Figs.\ \ref{specdttci}-(a)-(c).
Hence, the relaxation energy of the excited wave packet is 39meV, which
is consistent with the results of molecular orbital calculation shown
in Section \ref{MOcalc}.

\section{Discussion and conclusions}
\label{discussion}

In this paper we calculated the nonlinear optical response of wave packets
on quantized potential surfaces.
We found that, due to the discreteness of the vibrational eigenstates,
the probe signals as a function of pump-probe delay suddenly changes its
behavior following the stepwise change of the energy levels of the
molecule.
This aspect is particularly important when electron-vibration
interaction is weak, {\it i.e.}, the wave packet motion takes place in
the vicinity of the bottom of the excited-state PES.
In this case, a simple parabolic approximation for PESs is quite good and 
the analytical formula of the probe signal within impulse
approximation is sufficient to discuss experimental results.

The calculated results agree well with the pump-probe experimental results on DTTCI,
and it is shown that we can
estimate the strength of electron-vibration interaction directly from
experimental results.
These results are also consistent with the molecular orbital
calculations.
Thus, we stress that the accurate measurement of pump-probe
signals combined with theoretical calculations is a powerful tool
to understand the wave packet dynamics clearly.
Once the dynamics of the wave packets are sufficiently known, we try
to design the control method of them, which will bring us, {\it e.g.}, the
efficient way for chemical reaction control, optimization of
luminescence intensity from biological samples, or accurate control of qubits
stored in various materials such as quantum dots or molecules.
We mention that, as a typical example of the control method, the initial condition for the excited wave packet is
controlled by pulse chirp\cite{bardeen2,misawa,malkmus} and that the probe response is modulated by the presence of chirp.
We, however, also note that those results have not reached to control the wave packet motion
observing the quantum-mechanical wave packets and hence the
design of the control in coherent regime is still on the way even in
simple molecular systems.

We should mention that the present calculation does not take into account the
effect of decoherence of the vibrational states.
Since we are interested in molecules in solution, the relaxation
of excited wave packets is important to understand their physical
properties even in sub-picosecond regime.
These effects necessarily involve the mixed-state description of the
wave packets and coarse graining of the dynamical variables consisting
of the ``environment''.
It is necessary to obtain various knowledge of the total system
(molecules+environment) to develop a quantitative theory of the
decoherence in the present system, and thus the problem of decoherence
is
left for the future.
We, however, stress that the present results are still useful as a
basis of the study of wave packet dynamics even when the relaxation
processes are of importance.

\noindent {\bf acknowledgements}

The authors are grateful to K. Horikoshi for valuable discussions on
 pump-probe experiments of DTTCI.
K. I. also thanks S. Uchikoga and for helpful advice.

\appendix 

\section{Quantized form and the eigenstates of the Hamiltonian}
\label{eigen}

After canonical quantization of the Hamiltonian (\ref{ham0}), it is rewritten as
\begin{eqnarray}
{\cal H} & = & \hbar \sum_{i=1}^M \omega_i \left [a_i^\dagger a_i + \{ \varepsilon + {s_i^2 \over 1-t_i} + {1 \over 2}(1-\sqrt{1-t_i}) \right .\nonumber \\
& + & \left .  s_i (a_i^\dagger + a_i) - \frac{t_i}{4} (a_i^\dagger + a_i)^2\} |\uparrow \rangle \langle \uparrow | \right ] ,
\label{ham2}
\end{eqnarray}
where the creation operator of the vibration modes of the ground-state PES
is
\begin{equation}
a^\dagger_j = \sqrt{\frac{\omega}{2}} u_i  - i \frac{p_j}{\sqrt{2\omega}}.
\label{creation}
\end{equation}
It is diagonalized by introducing unitary transformation 
\begin{equation}
U = \prod_{i=1}^M U_i,
\end{equation}
where
\begin{equation}
U_i = \exp\left \{{\log(1-t_i) \over 8}((a_i^\dagger)^2-a_i^2) \right \}\exp\left \{{s_i \over 1-t_i}(a_i^\dagger-a_i) \right \}.
\end{equation}
The eigenvalue and the corresponding eigenvector of the Hamiltonian on
the excited-state PES are given by
\begin{eqnarray}
E^\uparrow_{n_1n_2,...n_M} & = & \hbar \left ( \sum_{i=1}^M n_i    \sqrt{1-t_i} \omega_i + \varepsilon\right ),\\
|n_1 n_2,... n_M \uparrow \rangle & = & \bigotimes_{i=1}^M U_i^\dagger|n_i \rangle,
\end{eqnarray}
and the creation operator of the phonon is
\begin{equation}
b_i^\dagger =a_i^\dagger \cosh \theta_i - a_i \sinh \theta_i - s_i(1-t_i)^{-3/4},
\end{equation}
where
\begin{equation}
e^{\theta_i} = (1-t_i)^{-1/4}.
\end{equation}
When the frequency of the vibrational modes for two PESs is the same,
{\it i.e.}, $t_i=0$, the above results are identical to those previously obtained\cite{old}.

The eigenstates on the excited-state PES for each mode $U_i^\dagger|n_i \rangle$ are expressed by
\begin{equation}
U_i^\dagger|n_i \rangle  =  {(b_i^\dagger)^{n_i} \over \sqrt{n_i!}} U_i^\dagger |0 \rangle 
 =  {(a_i^\dagger \cosh \theta_i - a_i \sinh_i \theta_i  - s_i(1-t_i)^{-3/4})^{n_i} \over \sqrt{n_i!\cosh \theta_i}}\exp \left \{{{\tanh \theta_i \over 2}(a_i^\dagger)^2} \right \}|0 \rangle,
\end{equation}
Thus, the matrix elements of the dipole operator $V$ is given by
\begin{equation}
\langle m_1 m_2 ... m_M \downarrow |V| n_1 n_2 ... n_M \rangle  = \prod_{i=1}^M \langle m_i |U_i^\dagger|n_i \rangle,
\end{equation}
where
\begin{eqnarray}
 \langle m_i |U_i^\dagger|n_i \rangle & = & \langle 0 | {a_i^{m_i}
   (b_i^\dagger)^{n_i}\over \sqrt{m_i!n_i!}} U_i^\dagger |0 \rangle
 \nonumber \\
& = & \langle 0 | {a_i^{m_i} (a_i^\dagger \cosh \theta_i  - a_i \sinh \theta_i  - s_i(1-t_i)^{-3/4})^{n_i} \over \sqrt{m_i!n_i!\cosh \theta_i}}\exp \left \{{{\tanh \theta_i \over 2}(a_i^\dagger)^2} \right \}|0 \rangle,\nonumber \\
& = & {1 \over \sqrt{n_i! m_i ! \cosh \theta_i}} \nonumber \\
& \times & \sum_{j = 0}^{n_i} \sum_{k
  =0}^{n_i-j}\sum_{l=0}^{\infty}{(-)^{n_i-j}\over 2^l l!} \cosh^{j-l} \theta_i \sinh^{k+l} \theta\i (s_i(1-t_i)^{-3/4})^{n_i-j-k} 
\langle 0 | a_i^{m_i} \{(a_i^\dagger)^j a_i^k \} (a_i^\dagger)^{2l}|0 \rangle \nonumber \\
& = & {1 \over \sqrt{n_i! m_i ! \cosh \theta_i}} \nonumber \\
& \times & \sum_{j = 0}^{n_i} \sum_{k =0}^{n_i-j}{(-)^{n_i-j} \over
  2^{m_i+k -j\over 2}({m_i+k-j \over 2})!}\cosh^{3j-m_i+k \over
  2}\theta_i \sinh^{m_i+3k-j \over 2} \theta_i
(s_i(1-t_i)^{-3/4})^{n_i-j-k} \nonumber \\
& \times &  \langle 0 |  a_i^{m_i} \{(a_i^\dagger)^j a_i^k \}
(a_i^\dagger)^{m_i+k-j}|0 \rangle.
\label{pol_t}
\end{eqnarray}
$\{(a_i^\dagger)^j a_i^k \}$ denotes the symmetrically ordered product of the
operator in the brace\cite{old},
and the sum over $k$ is taken only for the integral value of ${m_i+k-j \over 2}$.

\section{probe response by  ground-state wave packets}
\label{contr}

Amongst the third-order polarization terms which appear in the standard
perturbation theory\cite{mukamel}, the contribution of ground-state wave
packets to pump-probe signals is described by, for example,
\begin{equation}
P_g^{(3)}(t)  =  \int_0^t \! \! \int_0^t \! \!  \int_0^t dt_1 dt_2 dt_3 E_1(t_1)E_1(t-t_1-t_2)E_2(t-t_1-t_2-t_3)\langle 0 \downarrow |VV(t_1)V(t_1+t_2)V(t_1+t_2+t_3)| 0 \downarrow \rangle,
\label{gp}
\end{equation}
where $E_1(t)$ and $E_2(t)$ denote the electric field of the pump and
the probe pulse, respectively.

When the pump and the probe pulse are short and they are separated on
the temporal axis sufficiently, Eq.\ (\ref{gp}) for $M=1$ becomes
\begin{equation}
P_g^{(3)}(t) 
 \sim  \bar{S}_1^2 \bar{S}_2 
\sum_{n=0}^\infty e^{in\omega T} \langle se^{-i\omega \tau} ;\uparrow |V|n \downarrow
\rangle \langle n \downarrow |V e^{-\frac{i{\cal
      H}(t-T)}{\hbar}}V|0 \downarrow \rangle,\nonumber \\
\label{approxp}
\end{equation}
where $\bar{S}_1$ and $\bar{S_2}$ are the pulse area of the pump and
the probe pulses, respectively.
$\tau$ and $T$ denote the duration of the pump pulse and the
pump-probe delay.
Since $\langle 0 \downarrow |V e^{-\frac{i{\cal
      H}(t-T)}{\hbar}}V|0 \downarrow \rangle$ corresponds to the
ground state absorption which is independent of $T$, the vibronic
ground state $|0 \downarrow \rangle$ in Eq.\ (\ref{approxp})
does not contribute to the $T$-dependent part of $P_g^{(3)}(t)$,
and is irrelevant to the oscillatory behavior of the time-resolved spectra.
Hence, the magnitude of the temporal oscillation in time-resolved
spectra given by the ground-state wave packets is estimated by the factor
$|\langle se^{-i\omega \tau} ; \uparrow |V|n \downarrow  \rangle|^2$
which is interpreted as the transition probability of vertical transition between
excited-state wave packets and ground-state wave packets.

It is quite easy to extend the above discussion to multi-dimensional cases
($M \neq 1$), and we obtain the generalized formula to estimate the effect
of ground-state wave packets as
\begin{eqnarray}
r_{n_1 n_2 ... n_M}^j 
& = &  |\langle n_1 n_2 ... n_M \downarrow |V| s_1 e^{-i\omega_1 \tau} s_2
e^{-i\omega_2 \tau} ...s_M e^{-i\omega_M \tau} ;\uparrow  \rangle|^2
\nonumber \\
&  =  & \prod_{j=1}^M \exp \{-2s_j^2 (1-\cos \omega_j \tau)\} \frac{\{2s_j^2(1-\cos \omega_j \tau)\}^n_j}{n_j!}.
\label{trpr}
\end{eqnarray}

\end{document}